\documentclass[prd,nofootinbib,superscriptaddress,12pt]{revtex4-2}
 \usepackage{bm}
 \usepackage{amsmath}
 \usepackage{graphicx}
 \usepackage{float}
 \usepackage{amssymb}
 \usepackage{subfigure}
 \usepackage{amssymb}
 \usepackage{hyperref}
 \hypersetup{
 	colorlinks=true,
 	linkcolor=red,
 	citecolor=blue,
 }
 \usepackage[utf8]{inputenc}
 \hypersetup{
 	colorlinks=true,
 	linkcolor=red,
 	citecolor=blue,
 }\begin{document}
 	\title{Primordial black holes and secondary gravitational waves from chaotic inflation}

 	\author{Qing Gao}
    \email{gaoqing1024@swu.edu.cn}
    \affiliation{School of Physical Science and Technology, Southwest University, Chongqing 400715, China}

\begin{abstract}
Chaotic inflation is inconsistent with the observational constraint at 68\% CL.
Here, we show that the enhancement mechanism with
a peak function in the noncanonical kinetic term not only helps the chaotic model $V(\phi)=V_0\phi^{1/3}$ satisfy the observational constraint at large scales but also enhances the primordial scalar power spectrum by seven orders of magnitude at small scales.
The enhanced curvature perturbations can produce primordial black holes of different masses and secondary gravitational waves with different peak frequencies. We also show that the non-Gaussianities of curvature perturbations have little effect on the abundance of primordial black holes and energy density of the scalar-induced secondary gravitational waves.

\end{abstract}

\maketitle

\section{Introduction}	

The black holes detected by the Laser Interferometer Gravitational Wave Observatory (LIGO) Scientific Collaboration and the Virgo Collaboration
\cite{Abbott:2016blz,Abbott:2016nmj,Abbott:2017vtc,Abbott:2017oio,TheLIGOScientific:2017qsa,Abbott:2017gyy,LIGOScientific:2018mvr,Abbott:2020uma,LIGOScientific:2020stg,Abbott:2020khf,Abbott:2020tfl,Abbott:2020niy}
may be primordial black hole (PBH) dark matter (DM) \cite{Bird:2016dcv,Sasaki:2016jop}.
PBHs are formed during radiation domination by gravitational collapse when the density contrast of overdense regions  exceeds the threshold value at the horizon reentry  \cite{Carr:1974nx,Hawking:1971ei}, and they
 may constitute some or all DM
\cite{Ivanov:1994pa,Frampton:2010sw,Belotsky:2014kca,Khlopov:2004sc,Clesse:2015wea,Carr:2016drx,Inomata:2017okj,Garcia-Bellido:2017fdg,Kovetz:2017rvv,Carr:2020xqk}
and even explain the planet 9 \cite{Scholtz:2019csj}.
To produce PBHs, large density contrast is required, and the seeds of overdense regions can come from large
primordial curvature perturbations during inflation.
When large curvature perturbations reenter the horizon,
they become the sources of secondary gravitational waves (GWs)
due to the mixing of the tensor and curvature perturbations at the second order of perturbation \cite{Matarrese:1997ay,Mollerach:2003nq,Ananda:2006af,Baumann:2007zm}.
Therefore, in addition to PBHs, scalar-induced secondary gravitational waves (SIGWs) are generated  \cite{Matarrese:1997ay,Mollerach:2003nq,Ananda:2006af,Baumann:2007zm,Garcia-Bellido:2017aan,Saito:2008jc,Saito:2009jt,Bugaev:2009zh,Bugaev:2010bb,Alabidi:2012ex,Orlofsky:2016vbd,Nakama:2016gzw,Inomata:2016rbd,Cheng:2018yyr,Cai:2018dig,Bartolo:2018rku,Bartolo:2018evs,Kohri:2018awv,Espinosa:2018eve,Cai:2019amo,Cai:2019elf,Cai:2019bmk,Cai:2020fnq,Domenech:2019quo,Domenech:2020kqm,Pi:2020otn}.
These SIGWs are a part of the  stochastic  background
that can be detected by pulsar timing arrays (PTA) \cite{Ferdman:2010xq,Hobbs:2009yy,McLaughlin:2013ira,Hobbs:2013aka,Moore:2014lga}
and space-borne GW observatories, such as laser interferometer space antenna (LISA) \cite{Danzmann:1997hm,Audley:2017drz}, Taiji \cite{Hu:2017mde} and TianQin  \cite{Luo:2015ght}. Thus, the observation of PBH DM and SIGWs can elucidate primordial curvature perturbations and probe the physics in the early universe.

To produce detectable SIGWs and abundant PBH DM, the amplitude of primordial curvature perturbations must be as large as  $0.01$ \cite{Lu:2019sti,Sato-Polito:2019hws}.
Since the observation of the cosmic microwave background (CMB) constrains the amplitude of the primordial power spectrum to $A_s=2.1\times 10^{-9}$ at the pivotal scale $k_*=0.05\ \mathrm{Mpc}^{-1}$ \cite{Akrami:2018odb}, large primordial curvature perturbations can be obtained only at small scales.
Hence, some special mechanisms are needed to enhance the primordial power spectrum at small scales \cite{Gong:2017qlj,Martin:2012pe,Motohashi:2014ppa,Garcia-Bellido:2017mdw,Germani:2017bcs,Motohashi:2017kbs,Ezquiaga:2017fvi,Bezrukov:2017dyv,Espinosa:2017sgp,Ballesteros:2018wlw,Sasaki:2018dmp,Kamenshchik:2018sig,Gao:2018pvq,Dalianis:2018frf,Dalianis:2019vit,Passaglia:2018ixg,Passaglia:2019ueo,Fu:2019ttf,Fu:2019vqc,Xu:2019bdp,Braglia:2020eai,Gundhi:2020zvb,Zhou:2020kkf}.
One mechanism is to introduce a noncanonical kinetic term with a peak function \cite{Lin:2020goi,Yi:2020kmq,Yi:2020cut}. With this mechanism, both sharp and broad peaks  of the primordial power spectrum at small scales can be generated \cite{Yi:2020kmq,Yi:2020cut},
and it has been reported that the mechanism is effective for chaotic inflation with the potentials $V(\phi)=V_0\phi^{2/5}$ and $V(\phi)=\lambda \phi^4/4$, T-model, and natural inflation \cite{Lin:2020goi,Yi:2020kmq,Yi:2020cut,Gao:2020tsa}.
Non-Gausssianities of curvature perturbations in the models also have negligible effects on the PBH abundance
and energy density of SIGWs \cite{Zhang:2020uek}.
In this study, we investigated the production of PBHs and SIGWs in chaotic inflation
with the potentials $V(\phi)=V_0\phi^{1/3}$.

This paper is organized as follows:
Section \ref{sec2} discusses the enhancement of the power spectrum.
We also discuss non-Gausssianities of curvature perturbations and the energy scale of reheating in this model.
Section \ref{sec3} discusses the production of PBH DM, and the generation of SIGWs is discussed in Section \ref{sec4}. The effects of non-Gausssianities on the production of PBH DM and SIGWs are also discussed.
The conclusion is drawn in Section \ref{sec5}.

\section{Enhancement of scalar power spectrum }
\label{sec2}
The action for k/G inflation with noncanonical kinetic term is \cite{Lin:2020goi,Yi:2020kmq}
\begin{equation}\label{act1}
  S=\int d x^4 \sqrt{-g}\left[\frac{1}{2}R+X+G(\phi)X-V(\phi)\right],
\end{equation}
where $X=-g_{\mu\nu}\nabla^{\mu}\phi\nabla^{\nu}\phi/2$. The noncanonical kinetic term can be derived from G or k inflation, and we consider $M^{-2}_\text{Pl}=8\pi G=1$.
In Brans--Dicke theory, the noncanonical kinetic term is $\omega X/\phi$. If the zero point is shifted to a finite point $\phi_p$ and a small number $w$ is added to avoid singularity at $\phi_p$, then the noncanonical kinetic term is $h w X/[(\phi-\phi_p)/M_\text{Pl} +w]$, with $hw\sim O(1)$.
Based on this,
to enhance the scalar power spectrum at small scales, we use the peak function \cite{Yi:2020kmq,Yi:2020cut}
\begin{equation}\label{gfuncng}
G(\phi)=\frac{h/w^q}{1+\left(|\phi-\phi_p|/{w}\right)^q},
\end{equation}
where the dimensionless parameters $h$ and $w$ control the height and width of the peak, respectively, and $\phi_p$ and $q$ determine the position and shape of the peak, respectively.
The peak function is based on the noncanonical kinetic term $X/\phi$ in Brans--Dicke theory.
Away from the peak $\phi_p$, the peak function $G(\phi)$ is chosen to be negligible. Thus, at low energies after inflation, the noncanonical kinetic term can be neglected, and the standard canonical kinetic term is recovered.
With the noncanonical kinetic term, the background equations are
\begin{gather}
\label{Eq:eom1}
3H^2=\frac{1}{2}\dot{\phi}^2+V(\phi)+\frac{1}{2}\dot{\phi}^2G(\phi),\\
\label{Eq:eom2}
\dot{H}=-\frac{1}{2}[1+G(\phi)]\dot{\phi}^2,\\
\label{Eq:eom3}
\ddot{\phi}+3H\dot{\phi}+\frac{V_{\phi}+\dot{\phi}^2G_{\phi}/2}{1+G(\phi)}=0,
\end{gather}
where $G_\phi=dG(\phi)/d\phi$, $V_\phi=dV/d\phi$.
The curvature perturbation $\zeta$ satisfies the following equation:
\begin{equation}\label{zeta:k}
  \frac{d ^2 u_k}{d\eta^2}+\left(k^2-\frac{1}{z}\frac{d z^2}{d\eta^2}\right)u_k=0,
\end{equation}
where the conformal time $\eta=\int dt/a(t)$, $u_k=z\zeta_k$,  $z=a\dot{\phi}[1+G(\phi)]^{1/2}/H$, and $\zeta_k(t)$ is the Fourier transform of the curvature perturbation $\zeta(\vec{x},t)$.
In the slow-roll approximation $|\epsilon_i|\ll 1$
with \cite{Lin:2020goi}
\begin{equation}
\label{slrcond1}
\epsilon_1=-\frac{\dot{H}}{H^2},\quad \epsilon_2=-\frac{\ddot{\phi}}{H\dot{\phi}},\quad \epsilon_3=\frac{G_\phi\dot{\phi}^2}{V_\phi},
\end{equation}
the power spectrum of the curvature perturbation is \cite{Lin:2020goi}
\begin{equation}
\label{ps}
\begin{split}
\mathcal{P}_\zeta&=\frac{k^3}{2\pi^2}\left|\zeta_k\right|^2\\
&=\frac{H^4}{4\pi^2\dot{\phi}^2[1+G(\phi)]}\\
&\approx \frac{V^3}{12\pi^2V_{\phi}^2}[1+G(\phi)].
\end{split}
\end{equation}
According to Eq. \eqref{ps}, if the noncanonical function $G(\phi)$ is large enough,
the scalar power spectrum can be enhanced,
and this could be achieved using the peak function \eqref{gfuncng}.
When $\phi$ is around the peak $\phi_p$, which is chosen to correspond to small scales, $G(\phi_p)\approx h$.
Therefore, to enhance the power spectrum by seven orders of magnitude, the value of $h$ should be at least $10^7$. To make the peak function negligible away from the peak, $w$ must be small enough.
By contrast,
the peak function $G(\phi)$ also increases the number of $e$-folds $N$ before the end of inflation,
\begin{equation}\label{efold0}
N=\int_{\phi_e}^{\phi_*}\frac{V}{V_{\phi}}d \phi+\frac{V(\phi_p)}{V_{\phi}(\phi_p)}\int_{\phi_p+\Delta \phi}^{\phi_p-\Delta \phi}G(\phi) d\phi,
\end{equation}
where $\phi_*$ is the value of the scalar field
at the horizon exit where the pivotal scale $k_*$ leaves the horizon and $\phi_e$ is that at the end of inflation.
The first term in Eq. \eqref{efold0} is the contribution from the canonical scalar field, designated as $N_{eff}$, and the second term comes from the peak function.
To enhance the scalar power spectrum by seven orders of magnitude,
the contribution from the peak function to $N$ should be approximately $20$, and the remaining number of $e$-folds from the canonical scalar field is $N_{eff}\sim 40$.
As mentioned above, away from the peak, the peak function $G(\phi)$ is negligible and the usual inflation by canonical scalar field is recovered.
This implies that the peak function plays a role only on the peak $\phi_p$. It changes the number of $e$-folds effectively from $N_{eff}\sim 60$ to $N_{eff}\sim 40$ and enhances the scalar power spectrum around the peak. In addition, the effect of the peak function can be neglected, and the usual inflation from the canonical scalar field is considered.
For the same $\phi_e$, $\phi_*$ should be close to $\phi_e$ since $N_{eff}\sim 40$.Therefore, inflationary models that satisfy the CMB constraints with $N_{eff}=60$ would conflict with the observations.
For example, for chaotic inflation with the power-law potential \cite{Linde:1983gd,Lin:2015fqa},
\begin{equation}\label{pngb}
 V(\phi)=V_0\phi^n,
\end{equation}
with the slow-roll approximation, the scalar spectral index $n_s$ and tensor-to-scalar ratio $r$ are
\begin{equation}\label{ns}
n_s=1-\frac{n+2}{2N},
\end{equation}
\begin{equation}\label{r}
r=\frac{4n}{N}.
\end{equation}
If $N=60$, chaotic inflation is excluded by the Planck 2018 constraints \cite{Akrami:2018odb}
\begin{equation}
\label{cmb:con}
n_s = 0.9649\pm 0.0042  ~(68\% \text{CL}) ,\quad  r_{0.05} < 0.06 ~(95\% \text{CL}),
\end{equation}
at the 68\% CL.
With the peak function, $N$ is replaced with $N_{eff}\sim 40$.
If $n=1/3$, from Eqs. \eqref{ns} and \eqref{r}, we get
$n_s\approx0.971$ and $r\approx0.033$, which are consistent with the Planck 2018 results.
From the above slow-roll analysis, we understand the enhancement mechanism and why it overcomes the difficulty of maintaining the total number of $e$-folds at $N\sim 60$ while enhancing the power spectrum by seven orders of magnitude at small scales and satisfying the CMB constraints at large scales simultaneously.
Notably, around the peak, the slow-roll results are not applicable.
Thus, both the background and perturbation equations
must be solved numerically.
The slow-roll analysis clarifies the mechanism.
Here, we choose the power-law potential
$V(\phi)=V_0\phi^{1/3}$ and $\phi_*=4.65$ at the horizon exit where the pivotal scale $k_*=0.05$ $\text{Mpc}^{-1}$ leaves the horizon.

\begin{figure}[htbp]
 \centering
 \includegraphics[width=0.6\columnwidth]{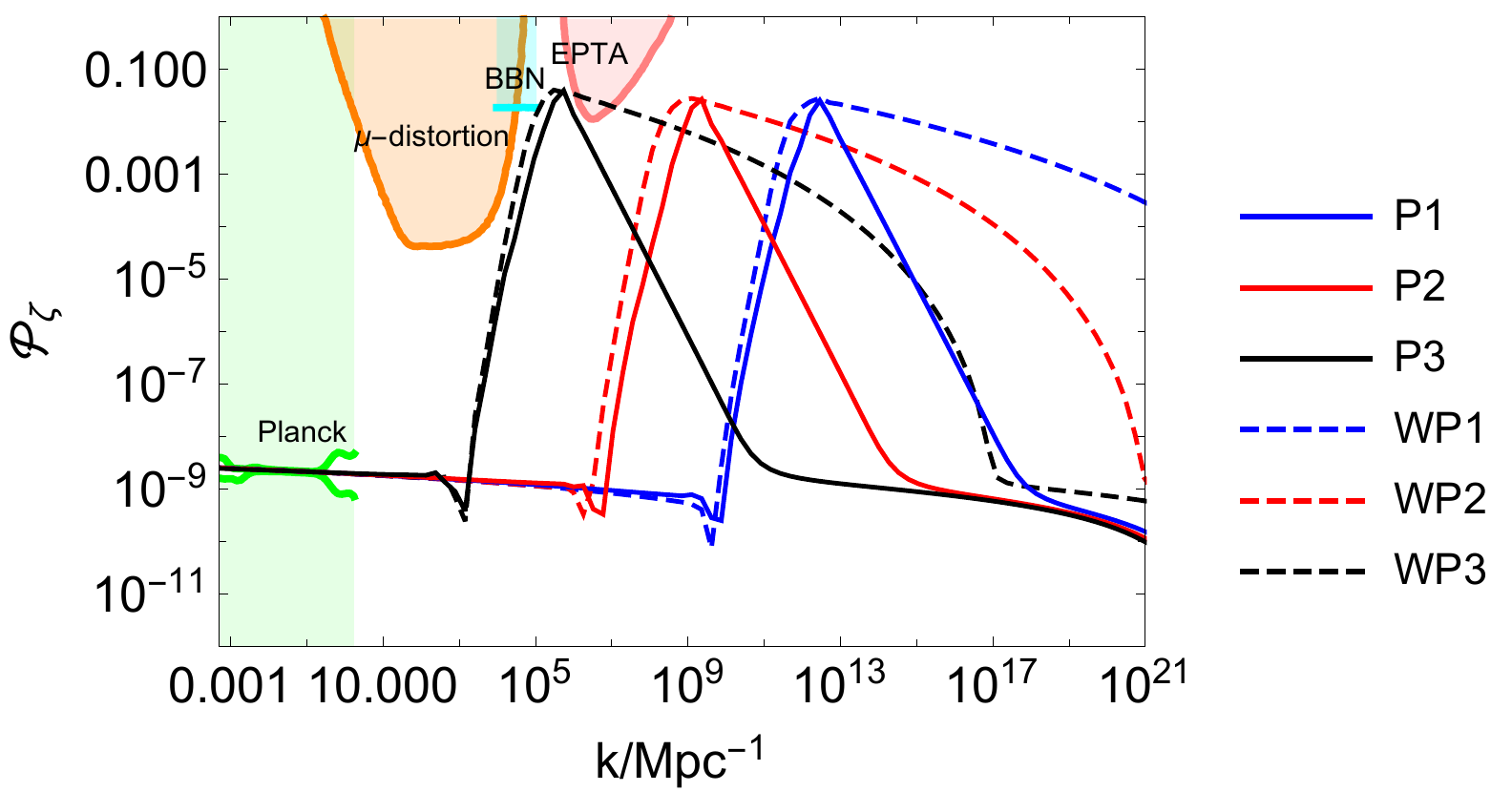}
 \caption{Results for the scalar power spectrum.
The solid lines denote models with the parameter $q=1$,
and the dashed lines denote models with the parameter $q=5/4$.
The blue, red, and black lines indicate models with peaks around
$10^{12}$, $10^{8}$, and $10^{5}$ Mpc$^{-1}$, respectively.
The parameters for the models and the peak scales $k_{\text{peak}}$ are shown in Table \ref{table1}.
The peak values of the power spectra are shown in Table \ref{table2}.
The light green-shaded region is excluded by CMB observations \cite{Akrami:2018odb}. The pink, cyan, and orange regions indicate the
constraints from PTA observations \cite{Inomata:2018epa},
the effect on the ratio between the neutron and proton
during the big bang nucleosynthesis (BBN) \cite{Inomata:2016uip},
and $\mu$-distortion of CMB \cite{Fixsen:1996nj}, respectively.}\label{pr}
\end{figure}

Choosing the energy scale $V_0$ for the potential so that the amplitude of the power spectrum at the pivotal scale is $A_s=2.1\times 10^{-9}$, $w=1.0\times 10^{-11}$
and the parameters $h$ and $\phi_p$ for the peak function, as shown in Table \ref{table1}, we numerically solve the background Eqs. \eqref{Eq:eom1}--\eqref{Eq:eom3}
and perturbation Eq. \eqref{zeta:k} to obtain the scalar power spectrum, as shown in Fig. \ref{pr}. In Table \ref{table1}, the label "P" indicates the models with the parameter $q=1$, which produce sharp peaks at small scales in the scalar power spectrum,
and "WP" indicates the models with the parameter $q=5/4$, which produce broad peaks at small scales in the scalar power spectrum. For these models,
the numerical results for $\phi_e$, number of $e$-folds, scalar spectral tilt $n_s$, and tensor-to-scalar ratio $r$ at the horizon exit are listed in Table \ref{table1}.
The numerical results show that the model with these parameter sets is consistent with the Planck 2018 results.
To distinguish various parameter sets in Table \ref{table1},
we use labels 1, 2, and 3 to represent the models with different peak scale $k_{\text{peak}}$ at which the scalar power spectrum has the maximum value.
The models with a peak scale around $10^{12}$, $10^{9}$,
and $10^{5}$ Mpc$^{-1}$ are labeled as 1, 2, and 3, respectively.
The numerical results of the peak scales $k_{\text{peak}}$ are also shown in Table \ref{table1}, and the peak values of the power spectra are shown in Table \ref{table2}.

\begin{table*}[htbp]
\begin{center}
	\renewcommand\tabcolsep{4.0pt}
	\begin{tabular}{cccccccccc}
		\hline
		\hline
		Model \quad & $V_0/10^{-10}$  &$h/10^{-2}$  &$\phi_p$ &$\phi_e$ &$N$&$n_s$&$r$&$k_{\text{peak}}/\text{Mpc}^{-1}$\\
		\hline
 		P1 \quad & 7.396   &$ 7.835$ & 2.46~ & 0.056 & 54.48~&$0.967$~&$0.039$~& $3.24\times 10^{12}$\\
        P2 \quad  & 7.354  & $5.890$& 3.24 & 0.057 & 53.68 &$0.968$&$0.039$&$2.13\times 10^{9}$\\
        P3 \quad  & 7.162 & $ 4.810$ & 3.96 & 0.057 &$53.32$&$0.972$&$0.038$&$ 5.47\times 10^{5}$\\
        WP1 \quad  & 7.628  & $ 1.625$& 1.73 & 0.057 & 69.93 &$0.965$&$0.04$&$ 2.32\times 10^{12}$\\
        WP2 \quad  & 7.629  &$ 0.8857$& 2.83 & 0.057 & 63.78 &$0.965$&$0.04$&$9.56\times 10^{8}$\\
        WP3 \quad & 7.613  &$ 0.6544$&3.61 & 0.057 & 62.15 &$0.965$&$0.04$&$ 5.36\times 10^{5}$\\
		\hline
		\hline
	\end{tabular}
	\caption{Model parameters and the numerical results.}
	\label{table1}
	\end{center}
\end{table*}

\begin{figure}
\centering
\includegraphics[width=0.46\textwidth]{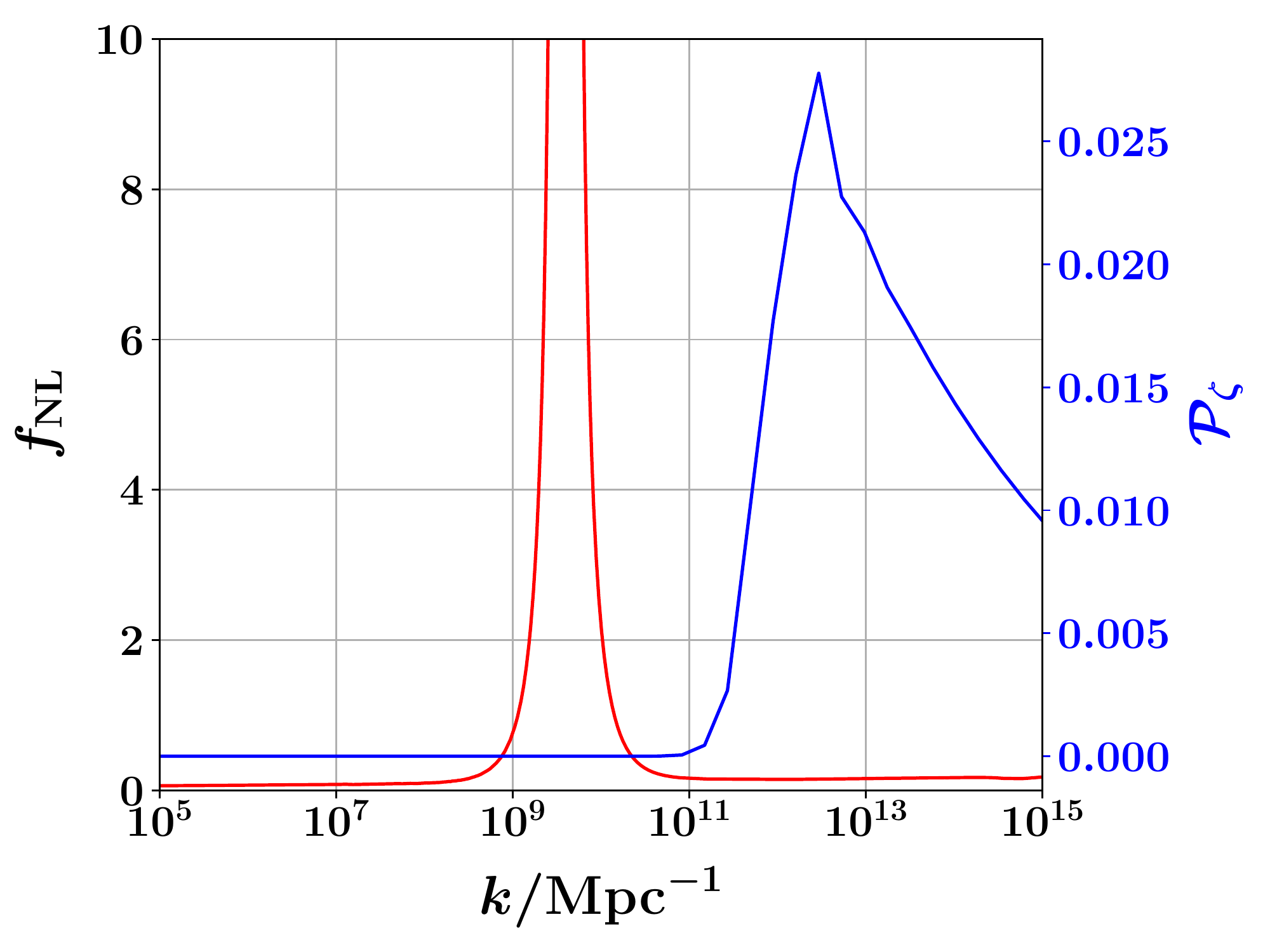}
\includegraphics[width=0.46\textwidth]{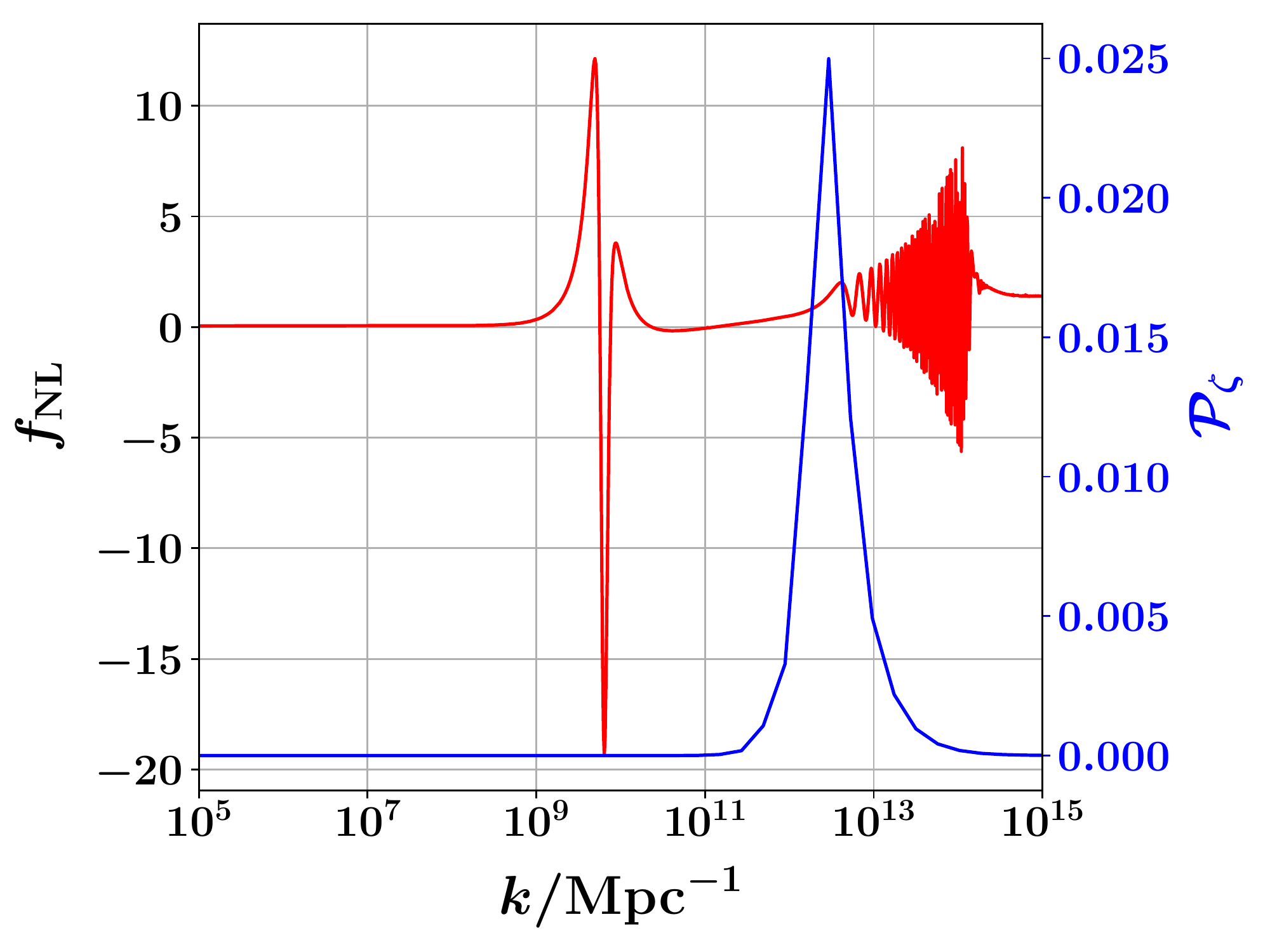}

\caption{Results of the non-Gaussianity parameters $f_{\text{NL}}$ (red lines) in the equilateral limit along with the primordial scalar power spectrum $\mathcal{P}_{\zeta}$  (blue lines) for models P1 and WP1.
  The left panel shows the results of model WP1, and the right panel shows those of model P1.}\label{ngeq}
\end{figure}

\begin{figure}[htbp]
  \centering
  \includegraphics[width=0.5\columnwidth]{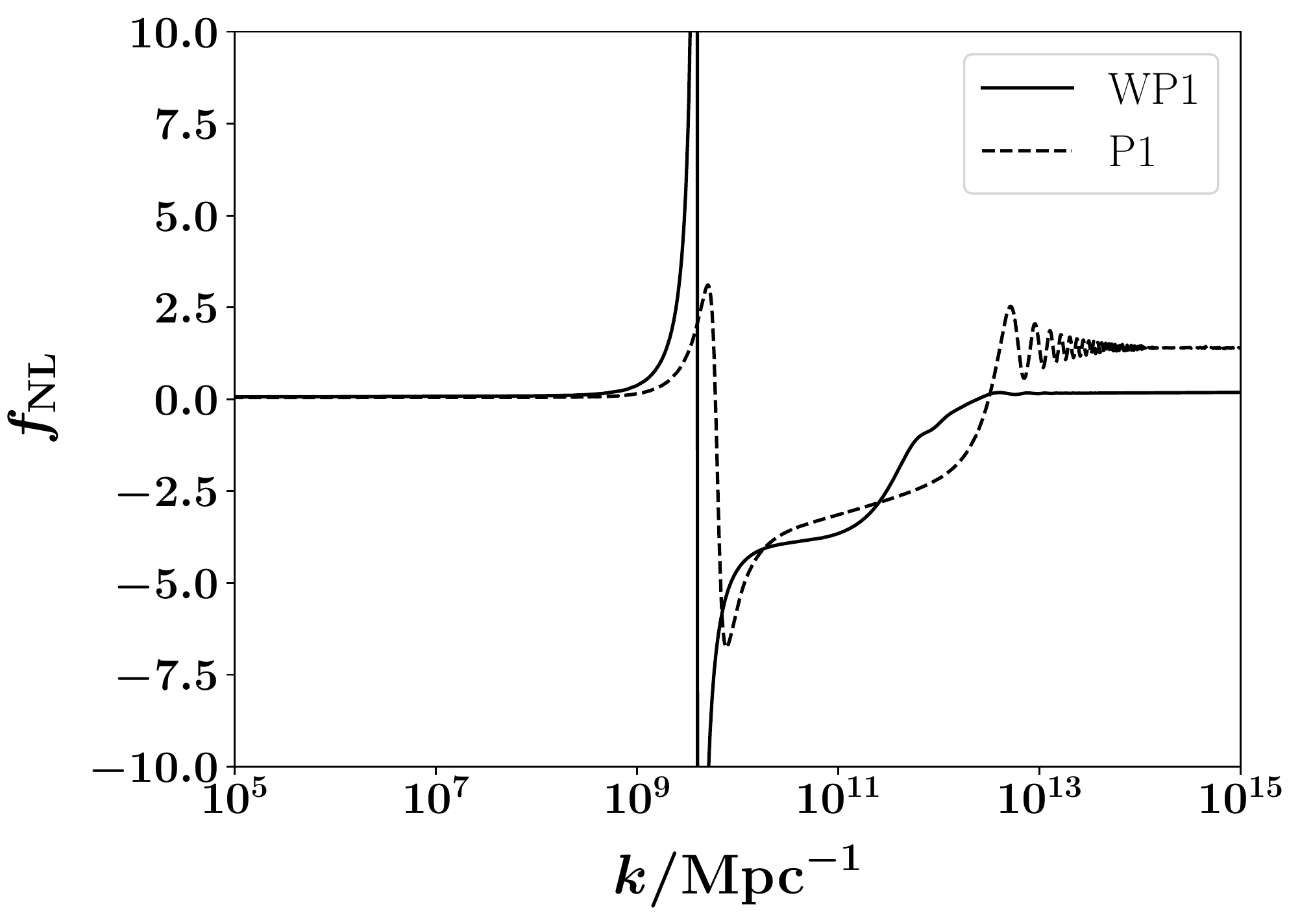}
  \caption{Results of the non-Gaussianity parameters $f_{\text{NL}}$ in the squeezed limit for models P1 (dashed line) and WP1 (solid line).}\label{ngsq}
\end{figure}

In addition to the two-point correlation function, we calculate the three-point correlation function to obtain the bispectrum $B_{\zeta}$ \cite{Byrnes:2010ft,Ade:2015ava},
\begin{equation}\label{Bi}
\left\langle\hat{\zeta}_{\bm{k}_{1}}\hat{\zeta}_{\bm{k}_{2}}\hat{\zeta}_{\bm{k}_{3}}\right\rangle=(2 \pi)^{3} \delta^{3}\left(\bm{k}_{1}+\bm{k}_{2}+\bm{k}_{3}\right) B_{\zeta}\left(k_{1}, k_{2}, k_{3}\right),
\end{equation}
where $\hat{\zeta}_{\bm{k}}$ is the corresponding quantum operator of the curvature perturbation $\zeta_k$.
The non-Gaussianity parameter $f_\text{NL}$ is \cite{Creminelli:2006rz,Byrnes:2010ft}
\begin{equation}\label{Fnl}
f_{\text{NL}} (k_1,k_2,k_3)=\frac{5}{6}\frac{B_{\zeta}(k_1,k_2,k_3)}{P_{\zeta}(k_1)
P_{\zeta}(k_2)+P_{\zeta}(k_2)P_{\zeta}(k_3)+P_{\zeta}(k_3)P_{\zeta}(k_1)},
\end{equation}
where $P_{\zeta}(k)=|\zeta_k|^2$. Using the numerical solution of the curvature perturbation, we calculate the  non-Gaussianity parameter $f_\text{NL}$ in the equilateral and squeezed limits and the results are shown in Figs. \ref{ngeq} and \ref{ngsq}, respectively.

The energy scale of reheating can also be estimated using this model.
For simplicity, we assume matter domination during reheating; thus, we have
\begin{equation}
\label{reht}
N=60.86-\ln \tilde{h}-\ln\frac{k}{a_0 H_0}-\frac{1}{3}\ln\frac{V_{e}^{1/4}}{\rho_{reh}^{1/4}}
+\ln\frac{V_*^{1/4}}{V_{e}^{1/4}}-\ln\left(\frac{10^{16}{\rm Gev}}{V_{*}^{1/4}}\right),
\end{equation}
where the Hubble constant $H_0=100\tilde{h}\ \text{km/s/Mpc}=67.4$ km/s/Mpc \cite{Aghanim:2018eyx}, $V_*$ and $V_e$ are the values of the potential at the horizon exit and end of inflation, respectively. Taking model P1 as an example, plugging the numerical results from Table \ref{table1} into Eq. \eqref{reht},
we obtain the energy scale of reheating $\rho^{1/4}_{reh}=7.4\times 10^{-6}$, which is around $10^{13}$ Gev.

\section{Primordial black hole dark matter}
\label{sec3}

The enhanced primordial curvature perturbations at small scales may form PBHs through gravitational collapse when they reenter the horizon during radiation domination.
The current fractional energy density of PBHs with respect to DM is \cite{Carr:2016drx,Gong:2017qlj}
\begin{equation}
\label{fpbheq1}
\begin{split}
Y_{\text{PBH}}(M)=&\frac{\beta(M)}{3.94\times10^{-9}}\left(\frac{\gamma}{0.2}\right)^{1/2}
\left(\frac{g_*}{10.75}\right)^{-1/4}\\
&\times \left(\frac{0.12}{\Omega_{\text{DM}}h^2}\right)
\left(\frac{M}{M_\odot}\right)^{-1/2},
\end{split}
\end{equation}
where $M$ is the mass of the PBH, $M_{\odot}$ is the solar mass, $\gamma= 0.2$ \cite{Carr:1975qj}, the current
energy density parameter of DM is $\Omega_{\text{DM}}h^2=0.12$ \cite{Aghanim:2018eyx},
the effective degrees of freedom $g_*=107.5$ for $T>300$ GeV,
and $g_*=10.75$ for $0.5\ \text{MeV}<T<300\ \text{GeV}$.
The fractional energy density of PBHs
at the formation can be approximated as \cite{Young:2014ana, Ozsoy:2018flq,Tada:2019amh}
\begin{equation}
\label{eq:beta}
\beta(M) \approx \sqrt{\frac{2}{\pi}}\frac{\sqrt{\mathcal{P}_{\zeta}}}{\mu_c}
\exp\left(-\frac{\mu_c^2}{2\mathcal{P}_{\zeta}}\right),
\end{equation}
$\mu_c=9\sqrt{2}\delta_c/4$ and
the critical density perturbation for the PBH formation is taken as $\delta_c=0.4$ \cite{Musco:2012au,Harada:2013epa,Tada:2019amh,Escriva:2019phb,Yoo:2020lmg}.
Some assumptions need to be made on $\mathcal{P}_\zeta$ to derive  the simple relationship \eqref{eq:beta} between $\beta(M)$ and $\mathcal{P}_\zeta$. Here, we use the simple relationship \eqref{eq:beta} to estimate the PBH abundance. Discussions on the dependence of PBH abundance on the detail of the curvature perturbation and statistics are available in Refs. \cite{Atal:2018neu,Germani:2018jgr,Germani:2019zez,Musco:2020jjb}.
The mass $M_\text{PBH}$ of PBH is related with the scale $K$ as
\begin{equation}\label{mass:pm}
M_\text{PBH}=3.68\left(\frac{\gamma}{0.2}\right) \left(\frac{g_*}{10.75}\right)^{-1/6} \left(\frac{k}{10^6\ \text{Mpc}^{-1}}\right)^{-2} M_\odot.
\end{equation}
Therefore, the peak mass $M_\text{peak}$ of PBH can be determined from the peak scale $k_\text{peak}$ of the primordial power spectrum using the above relation \eqref{mass:pm}.

\begin{figure}[htbp]
  \centering
  \includegraphics[width=0.6\columnwidth]{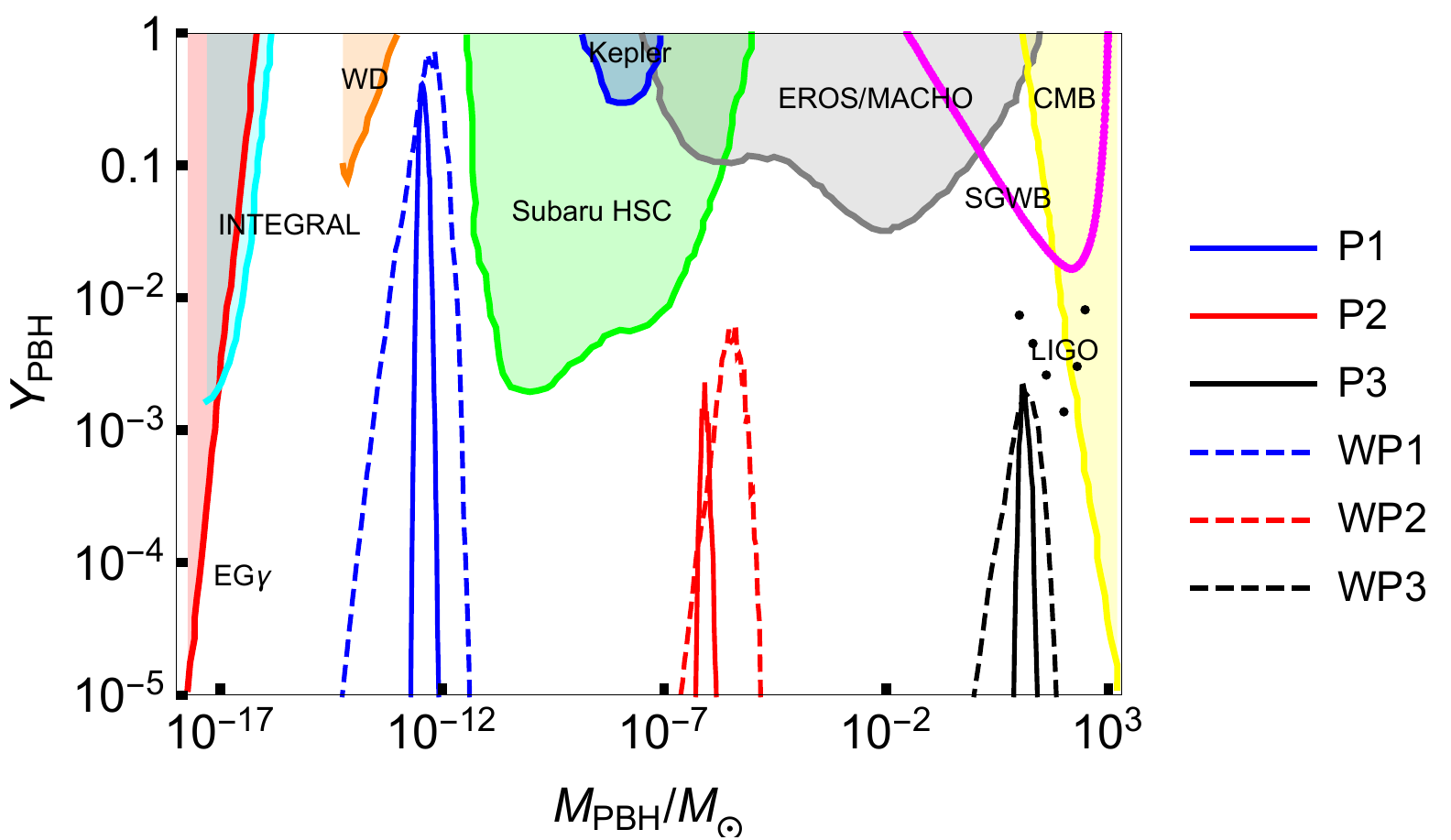}
  \caption{Results for PBH abundances.
  The parameters for the models are shown in Table \ref{table1},
 and the peak abundance and peak mass of PBHs are shown in Table \ref{table2}.
The shaded regions show the observational constraints on the PBH abundance:
the yellow region is the accretion constraints by CMB \cite{Ali-Haimoud:2016mbv,Poulin:2017bwe};
the red region is from extragalactic gamma-rays by PBH evaporation \cite{Carr:2009jm} (EG$\gamma$); the cyan region is from galactic center 511 keV gamma-ray line (INTEGRAL) \cite{Laha:2019ssq,Dasgupta:2019cae};
the orange region is from white dwarf explosion (WD) \cite{Graham:2015apa};
the green region is from microlensing events with Subaru HSC \cite{Niikura:2017zjd};
the blue region is from the Kepler satellite \cite{Griest:2013esa}; and
the gray region is from the EROS/MACHO \cite{Tisserand:2006zx}.
The solid magenta line indicates the constraints on the stochastic gravitational wave background by LIGO \cite{Raidal:2017mfl}, and the black dots show the limits from the LIGO merger rate \cite{Ali-Haimoud:2017rtz}.}\label{fpbh}
\end{figure}

Combining the numerical results
of the power spectra in Fig. \ref{pr} and Eqs. \eqref{fpbheq1} and \eqref{eq:beta},
we get the PHB abundance and peak masses of PBHs, as shown in Fig. \ref{fpbh} and Table \ref{table2}.
The results show that different masses of PBHs correspond to
different peak scales in the scalar power spectrum.
Taking different values of $\phi_p$, PBHs with masses of approximately $10^{-13}\ M_{\odot}$,
$10^{-6}\ M_{\odot}$, and $10\ M_{\odot}$ can be generated.
The PBHs observed by LIGO/Virgo collaboration could be explained by the stellar-mass PBHs.
The peak abundance $Y_\text{PBH}^\text{peak}$ of PBHs with a mass of approximately $10^{-13}\ M_{\odot}$ is almost 1. Therefore, they can
make up almost all DM.
PBH DM with a mass of approximately $\mathcal{O}(1)M_{\oplus}$ may explain planet 9.

\begin{table*}[htbp]
\begin{center}
	\renewcommand\tabcolsep{4.0pt}
	\begin{tabular}{ccccc}
		\hline
		\hline
		Model \quad   &$\mathcal{P}_{\zeta(\text{peak})}$& $M_\text{peak}/M_\odot$&$Y_\text{PBH}^\text{peak}$& $f_c/\text{Hz}$\\
		\hline
 		P1 \quad   &$0.0251$&$3.51\times10^{-13}$&$0.411$&$4.54\times 10^{-3}$\\
        P2 \quad   &$0.0269$&$8.13\times10^{-7}$&$2.30\times 10^{-3}$&$3.59\times 10^{-6}$\\
        P3 \quad   &$0.0366$&$12.30$&$2.16\times 10^{-3}$&$8.53\times 10^{-10}$\\
        WP1 \quad   &$0.0258$&$6.83\times10^{-13}$&$0.696$&$4.41\times 10^{-3}$\\
        WP2 \quad   &$0.0285$&$4.03\times 10^{-6}$&$6.37\times10^{-3}$&$1.71\times10^{-6}$\\
        WP3 \quad   &$0.0366$&$12.80$&$1.88\times10^{-3}$&$6.05\times 10^{-10}$\\
		\hline
		\hline
	\end{tabular}
	\caption{Results for the peak values of the primordial scalar power spectra, peak mass, and peak abundance of PBH, and the peak frequency of SIGWs.
The parameters for the models are shown in Table \ref{table1}.}
\label{table2}
\end{center}
\end{table*}

Since the major contribution comes from the peak of the power spectrum,
for good approximation of the PBH abundance at the peak by considering non-Gaussianities is \cite{Saito:2008em}
\begin{equation}
\label{beta}
\beta(M_\text{peak})=\frac{1}{\sqrt{2\pi}}\int_{\tilde{\zeta}_{th}}[(\tilde{\zeta}^2-1)-(\tilde{\zeta}^5-8\tilde{\zeta}^3+9\tilde{\zeta})
\mathcal{J}_\mathrm{peak}]e^{-\tilde{\zeta}^2/2},
\end{equation}
where $\tilde{\zeta}=\zeta/\sqrt{\mathcal{P}_\zeta(k_\text{peak})}$ and
\begin{equation}
\label{jeq2}
\mathcal{J}_\mathrm{peak}=\frac{3}{20\pi}f_\mathrm{NL}(k_\mathrm{peak},k_\mathrm{peak},k_\mathrm{peak})\sqrt{\mathcal{P}_\zeta(k_\mathrm{peak})}.
\end{equation}
Therefore, the parameter $\mathcal{J}_\mathrm{peak}$ can be used to evaluate the effect of the non-Gaussianity of curvature perturbation $\zeta$ on the PBH abundance.
If $\mathcal{J}_\mathrm{peak}\ll 1$, then the effect of non-Gaussianities on the PBH abundance is negligible.
Fig. \ref{ngeq} shows that $f_\mathrm{NL}(k_\mathrm{peak},k_\mathrm{peak},k_\mathrm{peak}) \sim \mathcal{O}(1)$ and $\mathcal{P}_\zeta (k_\mathrm{peak})\sim \mathcal{O}(0.01)$.
Plugging these numbers into Eq. \eqref{jeq2}, we obtain
$\mathcal{J}_\mathrm{peak} \ll 1$.
Hence, we infer that the effect of non-Gaussianity on PBH abundance is negligible, and we can neglect non-Gaussianity while calculating the PBH abundance in this model.
Even though $f_\text{NL}$ can be very large before the peak scale, at that scale, the amplitude of the primordial scalar power spectrum is very small.
Therefore, the parameter $\mathcal{J}$ is also small, and the effect of non-Gaussianity on PBH abundance is negligible in this model.

\section{Scalar-induced secondary gravitational waves}
\label{sec4}

During the production of PBHs, enhanced curvature perturbations at small scales induce secondary GWs due to the mixing of scalar and tensor perturbations at the second-order perturbation.
The current energy density of the SIGWs is
\begin{equation}\label{gwres2}
\Omega_{\text{GW}}(k,\eta_0)=\Omega_{\text{GW}}(k,\eta)\frac{\Omega_{r}(\eta_0)}
{\Omega_{r}(\eta)},
\end{equation}
where $\Omega_r$ is the fraction energy density of radiation.
The energy density of SIGWs
during the radiation domination is given by \cite{Inomata:2016rbd,Kohri:2018awv,Espinosa:2018eve}
\begin{equation}
\label{gwres1}
\begin{split}
\Omega_{\mathrm{GW}}(k,\eta)=&\frac{1}{6}\left(\frac{k}{aH}\right)^2
\int_{0}^{\infty}dv\int_{|1-v|}^{1+v}du\left\{ \left[\frac{4v^2-(1-u^2+v^2)^2}{4uv}\right]^2\right.\\
&\left. \times \overline{I_{\text{RD}}^{2}(u, v, x\to \infty)} \mathcal{P}_{\zeta}(kv)\mathcal{P}_{\zeta}(ku)\right\}.
\end{split}
\end{equation}
The time-averaged kernel function $\overline{I_{\text{RD}}^{2}}$ is given by \cite{Espinosa:2018eve,Lu:2019sti}

\begin{equation}
\label{irdeq}
\begin{split}
    \overline{I^2_{\text{RD}}(u,v,x\rightarrow\infty)}
    =&\frac{1}{2x^2}\left[\left(\frac{3\pi(u^2+v^2-3)^2\Theta(u+v-\sqrt3)}{4u^3v^3}
    +\frac{T_c(u,v,1)}{9}\right)^2 \right.\\
    &\left.\qquad +\left(\frac{\tilde{T}_s(u,v,1)}{9}\right)^2\right],
\end{split}
\end{equation}
$T_c(u,v,1)$ and $\tilde{T}_s(u,v,1)$ are given in Ref. \cite{Lu:2019sti}.

\begin{figure}[htbp]
  \centering
  \includegraphics[width=0.6\columnwidth]{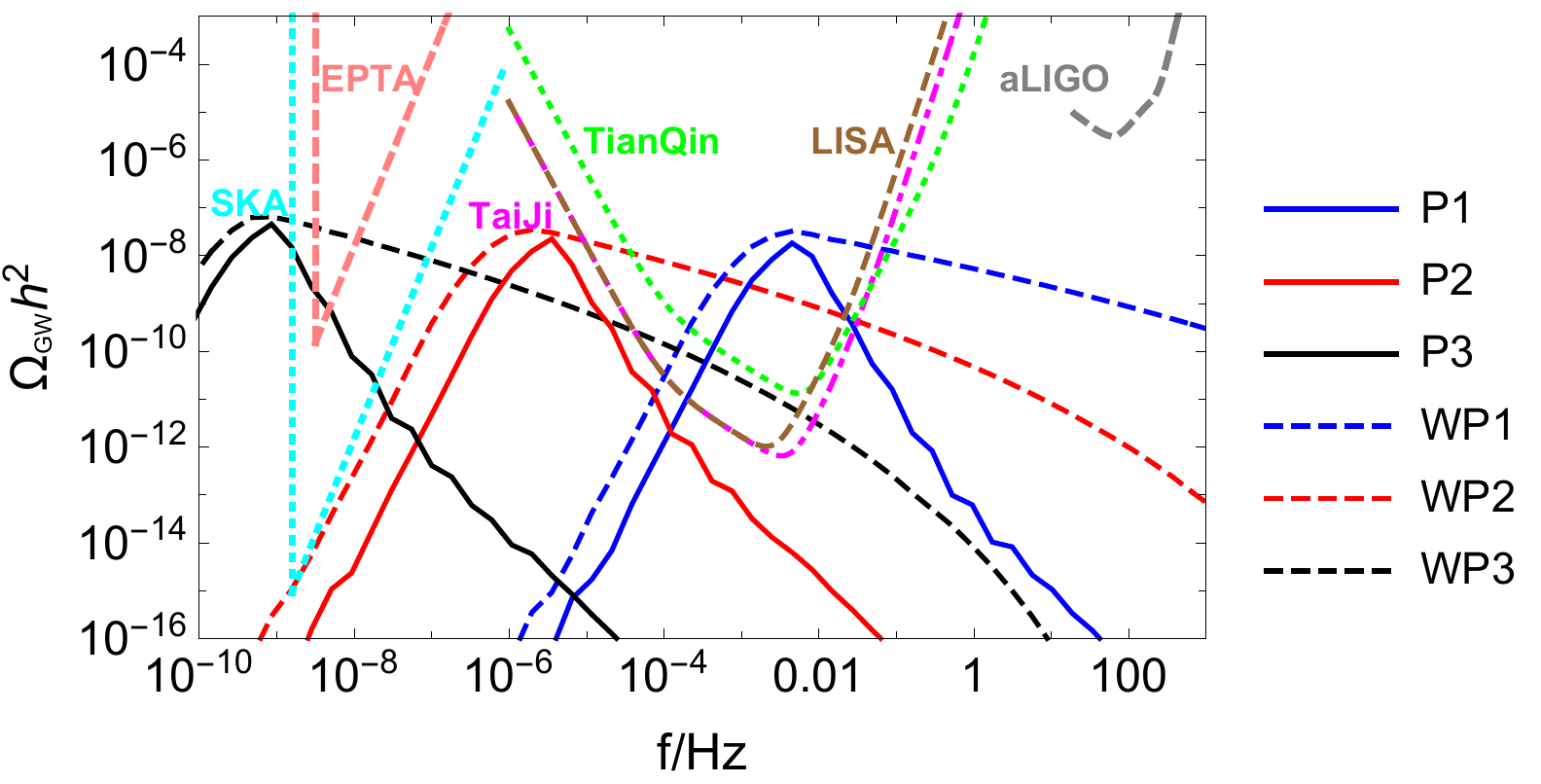}
  \caption{Energy densities of SIGWs.
  The parameters for the models are shown in Table \ref{table1},
  and the peak frequencies are shown in Table \ref{table2}.
The dashed pink curve denotes the EPTA limit \cite{Ferdman:2010xq,Hobbs:2009yy,McLaughlin:2013ira,Hobbs:2013aka} ,
the dotted cyan curve denotes the SKA limit \cite{Moore:2014lga},
the dashed green curve at the middle denotes the TianQin limit \cite{Luo:2015ght},
the dot--dashed magenta curve shows the TaiJi limit \cite{Hu:2017mde},
the dashed brown curve shows the LISA limit \cite{Audley:2017drz},
and the dashed gray curve denotes the aLIGO limit \cite{Harry:2010zz,TheLIGOScientific:2014jea}.}\label{gw}
\end{figure}

Substituting the numerical results of power spectra in Fig. \ref{pr} into  Eqs. \eqref{gwres2}, \eqref{gwres1} and \eqref{irdeq},
we obtain the current energy density of SIGWs,
and the result is shown in Fig. \ref{gw} and Table \ref{table2}.
The peak frequencies of the SIGWs are in the scale of mHz, $\mu$Hz, and nHz, respectively.
Fig. \ref{gw} shows that for models WP1, WP2, and WP3, $\Omega_{\text{GW}}$ has a broad shape that spans a wide frequency band, which is attributed to the broad peaks of the enhanced power spectrum in the models.
Models P3 and WP3 produce the stellar-mass PBHs, but model WP3 is
excluded by the EPTA data \cite{Ferdman:2010xq,Hobbs:2009yy,McLaughlin:2013ira,Hobbs:2013aka}.
Model P3 will be tested by SKA. Models P2 and WP2 produce the earth-mass PBHs, which can explain planet 9. The frequencies of SIGWs in model P2 lie in the $10^{-6}$ Hz band, and those in model WP2 span into the mHz band. Models P1 and WP1 can explain DM in terms of PBHs, and the frequencies of SIGWs produced in these models are in the mHz band. Hence, models WP2, P1, and WP1 will be tested by LISA/Taiji/TianQin.

To consider the effect of non-Gaussianities on SIGWs, we require $f_{\text{NL}}^2 \mathcal{P}_\zeta\gtrsim 1$ \cite{Cai:2018dig}. The numerical results shown in Figs. \ref{pr} and \ref{ngeq} reveal that the effect of non-Gaussianities on SIGWs is negligible.

\section{Conclusion}
\label{sec5}

Employing the peak function in the noncanonical kinetic term,
we show that the primordial curvature perturbations for chaotic inflation with the potential $V(\phi)=V_0\phi^{1/3}$ are not only consistent with the Planck 2018 observations at large scales
but also enhanced by seven orders of magnitude at small scales.
The enhanced power spectrum at small scales produces abundant PBHs and observable SIGWs.
Non-Gaussianities of the enhanced curvature perturbations have little effect on the PBH abundance and energy density of SIGWs.
By varying the model parameters, which we labeled as models 3, 2, and 1, the power spectrum is enhanced to the order of $0.01$ at the scales of $10^{5}$, $10^{9}$, and $10^{12}$ Mpc$^{-1}$, respectively.
The production of PBHs with the stellar mass,
earth mass, and mass of about  $10^{-13}\ M_{\odot}$
is accompanied by the generation of SIGWs with peak frequencies in the scale of nHz, $\mu$Hz, and mHz, respectively, when the enhanced scalar perturbations reenter the horizon during radiation domination.
PBHs with the mass of approximately $10^{-13}\ M_{\odot}$ produced in models P1 and WP1 can account for almost all DM,
The earth-mass PBHs produced in models P2 and WP2 may explain planet 9,
and the stellar-mass PBHs produced in models P3 and WP3 could be the black holes detected by LIGO and Virgo.
Model WP3 has broad peaks for both the power spectrum and SIGWs and is excluded by the EPTA data, and model P3 can be tested by SKA.
Models WP2, WP1, and P1  can be tested by LISA/Taiji/TianQin.

In conclusion, the enhancement mechanism by
a peak function in the noncanonical kinetic term
is effective for the chaotic potential $V(\phi)=V_0\phi^{1/3}$.

\begin{acknowledgments}
This work is supported by the National Key Research \& Development Program of China (Grant No. 2020YFC2201504)
and the Venture \& Innovation Support Program for Chongqing Overseas Returnees (Grant No. CX2020083).

\end{acknowledgments}


\end{document}